\documentclass[aps,showpacs,two column]{revtex4}
\usepackage{graphicx}% Include figure files
\usepackage{bm}% bold math
\usepackage{amsfonts,amsmath,amssymb}
\usepackage[squaren]{SIunits}

\begin{document}

\title{Intrinsic electron-doping in nominal ``non-doped" superconducting
(La,Y)$_2$CuO$_4$ thin films grown by dc magnetron sputtering}
\author{L. Zhao, G. Wu, R. H. Liu and X. H. Chen}
\altaffiliation{Corresponding author} \email{chenxh@ustc.edu.cn}
\affiliation{Hefei National Laboratory for Physical Science at
Microscale and Department of Physics, University of Science and
Technology of China, Hefei, Anhui 230026, People's Republic of
China}

\begin{abstract}

The superconducting nominal ``non-doped"
La$_{1.85}$Y$_{0.15}$CuO$_{4}$ (LYCO) thin films are successfully
prepared by dc magnetron-sputtering and \emph{in situ}
post-annealing in vacuum. The best T$_{\texttt{C0}}$ more than 13K
is achieved in the optimal LYCO films with highly pure c-axis
oriented T'-type structure.  In the normal state, the
quasi-quadratic temperature dependence of resistivity, the negative
Hall coefficient and effect of oxygen content in the films are quite
similar to the typical Ce-doped n-type cuprates, suggesting that
T'-LYCO  shows the electron-doping nature like known n-type
cuprates, and is not a band superconductor as proposed previously.
The charge carriers are considered to be induced by oxygen
deficiency.
\end{abstract}
\vskip 15 pt


\narrowtext

\maketitle

\section{INTRODUCTION}

Although the intense research since the discovery of
high-temperature superconductors by Bednorz and M\"{u}ller in
1986\cite{Bednorz}, the underlying mechanism for high temperature
superconductivity remains elusive. It is well known that the parent
compounds of all the high-temperature superconductors are
antiferromagnetic half-filled Mott insulators in which the strong
electronic correlation exists. Upon increasing doping holes or
electrons to CuO$_2$ sheets, the long-range antiferromagnetism is
destroyed and superconductivity can be induced at the certain doping
levels.

However, the new class of T'-structure cuprate superconductors,
recently discovered by NTT research group, are nominal ``non-doped"
T'-(La, RE)$_{2}$CuO$_{4}$ (RE = Sm, Eu, Gd, Tb, Lu, and Y). RE is
the isovalent cation with La and don't provide necessary charge
carriers to CuO$_2$ planes. They claimed that oxygen deficiencies
can't afford a main source of charge carriers and these new
superconductors are most plausibly so-called ``band superconductors"
\cite{naito, noda}. But they lacked further investigation on
physical properties of these new class of superconductors.
Furthermore, no other group can reproduce their work and acquire
superconducting samples. Only one unsuccessful attempt was reported
recently \cite{Idemoto}. It is urgent to confirm these results at
present.

According to our experimental experiences in the synthesis of
La-based 214 cuprate films such as La$_{2-x}$Ce$_{x}$CuO$_{4}$(LCCO)
films by sputtering and pulsed laser deposition (PLD) in the last
several years \cite{thesis}, we studied the La-Y-Cu-O system. In
this paper, we report the successful growth of the superconducting
T'-LYCO films by dc magnetron sputtering. The behaviors of
resistivity and Hall effect are investigated to disclose the nature
of charge carriers in T'-LYCO.

\section{EXPERIMENT DETAILS}

The La$_{1.85}$Y$_{0.15}$CuO$_{4}$ (LYCO) films were fabricated by
on-axis dc magnetron sputtering using a stoichiometric ceramic
target. The target was synthesized by conventional solid state
reaction. The appropriate powders of La$_2$O$_3$, Y$_2$O$_3$ and
CuO of high purity with the composition ratio of cation La: Y:
Cu=1.85:0.15:1.0, were mixed, grounded and sintered in air at
980$\celsius$ for 48 hours with several intermediate regrindings.
The calcined powder was then pressed into a 5 mm-thick disk with
50mm in diameter and was sintered at 1050$\celsius$ for 24 hours.

The (100)-cut SrTiO$_3$ (d=3.905\AA) and LSAT
[(LaAlO$_3$)$_{0.3}$(Sr$_2$AlTaO$_6$)$_{0.7}$] (d=3.868\AA) were
used as substrates for deposition of LYCO films. The substrates were
fixed to the heater which was located right above the target at a
height of 35mm. The base pressure of the chamber prior to deposition
was below $5\times10^{-4}$Pa. The sputtering gas was pure argon, and
the pressure in the chamber was kept at about 60Pa during the
deposition. Subsequently, LYCO was deposited at the temperature
T$_D$ = 600- 755$\celsius$ . After deposition, the samples were
first slowly cooled down to T$_A$ =600$\celsius$ in high vacuum
(below $10^{-4}$Pa), then annealed at this temperature for 15-30
minutes to remove the excess apical oxygen and finally followed by
very slow cooling to room temperature in high vacuum. usually 15
minute's annealing in high vacuum is enough and longer annealing
have no improving effect on the superconducting properties of the
LYCO films. During the deposition, the sputtering current was
ordinarily kept at about 200mA and the typical film thickness is
around 3000\AA ~after one hour's deposition. For these films thicker
than 2500\AA, there is no distinct difference on the transport
properties of these films grown on whether STO or LSAT substrates in
our experiments.

All the films are characterized by an x-ray diffractometer
(Panalytical, X'pert Pro), using Cu K$\alpha$ radiation. Their
surface morphology and thickness are observed using an atomic
force microscope (AFM, Digital Instruments). The measurements of
resistivity and Hall coefficients are carried through by standard
6-lead technology, using a  superconducting quantum interference
device (SQUID, Quantum Design, MPMS-7\emph{XL}). The Ag electrodes
were deposited by evaporation through a copper mask.

\section{RESULTS AND DISCUSSION}

Fig. 1 shows the x-ray diffraction pattern of the
La$_{1.85}$Y$_{0.15}$CuO$_{4}$ target. Except the trace of
un-reacted yttrium oxide, all the peaks are indexed assuming the
La$_2$CuO$_4$-like T-214 structure. The calculated lattice constants
of the orthorhombic lattice are, a = 5.355\AA , b = 5.399\AA ,and c
= 13.136\AA. These lattice constants are slight less than these of
La$_2$CuO$_4$ (a = 5.356\AA , b = 5.402\AA , c = 13.14\AA),
consistent with the partial substitution of larger La$^{3+}$ (ion
radii is 1.216\AA) by smaller Y{$^{3+}$ (1.075\AA).
\begin{figure}
\includegraphics[width=0.40\textwidth]{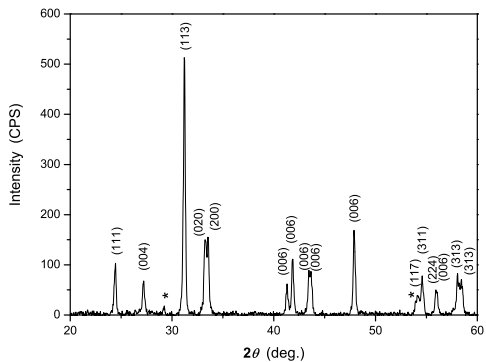}
\caption{ Powder x-ray diffraction patterns of La$_{1.85}$Y$_
{0.15}$CuO$_{4}$ target, the peak marked with stars is from a
small quantity of un-reacted Y$_2$O$_3$.\\}
\end{figure}

La$^{3+}$ is the largest ion among lanthanide series, and  the
detailed analysis on the perovskite crystallographic Goldshimidt
tolerance factor \emph{t}, have told us that La-214 tends to form
the T-type phase while T'-phase is unstable at relative lower
synthesis temperature \cite{goodenough}. By extrapolating the T/T'
phase boundary in the La$_{2-y}$Nd$_{y}$CuO$_{4}$ system to y=0,
Manthiram and Goodenough\cite{goodenough} predicted that
T'-La$_2$CuO$_4$ can only be stabilized below 425$\celsius$.
Although partially substitution of La$^{3+}$ by smaller Y$^{3+}$
will reduce slightly the average ion radius of A-site, and shift
slightly the T/T' phase boundary to higher temperature, It is still
too hard to realize the preparation of T'-phase bulk material.
Conventional solid state reaction can only achieve T-phase.
Nevertheless compared with the solid-state reaction process,  the
preparation of thin films can usually be realized at relative lower
temperature. Furthermore the appropriate epitaxy strain through
suitable substrates can stabilize some meta-stable phases
\cite{Tsukada}.

The x-ray diffraction patterns of LYCO films deposited at various
temperatures are shown in Fig. 2. The c-axis lattice constant of the
LYCO is 12.45\AA~ for T'-type structure and 13.136\AA~ for T-type
structure, respectively. Therefore, these two kinds of competing
structures can easily be distinguished in x-ray diffraction. As
mentioned above, the phase competition between the T'- and T-
structures also exists during the growth of the LYCO films and the
T'-phase is preferable at lower deposition temperature.

\begin{figure}
\includegraphics[width=0.4\textwidth]{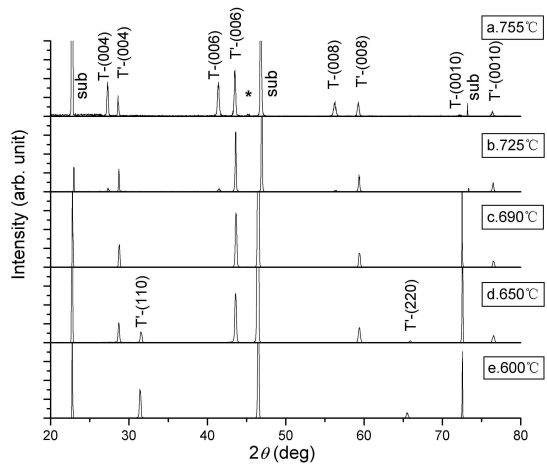}
\caption{ X-ray diffraction patterns($\theta-2\theta$ scan) of LYCO
films grown on STO and LSAT substrates at various deposition
temperatures 755$\celsius$, 725$\celsius$, 690$\celsius$,
650$\celsius$, and 600$\celsius$. The peak marked with ``$\ast$" is
the (200) reflection belonging to Cu$_2$O impurity. The un-indexed
peaks are due to the substrates.}
\end{figure}

From Fig. 2, one can see that the LYCO films with pure T'-phase can
be synthesized only when the deposition temperature T$_{D}$ is near
690$\celsius$. Upon increasing T$_{D}$, the diffraction peaks
belonging to T-phase develop, indicating  the coexistence of  two
phases in the films. They are both highly c-axis oriented and only
(00\emph{l}) peaks exist, except the small (200) peak of Cu$_2$O
impurity (marked with ``$\ast$" in Fig.2, appears only when
deposition temperature is higher than 725$\celsius$). With
increasing T$_{D}$, the proportion of T'-phase component in the
films decreases rapidly, and that of T-phase component increases
correspondingly. It is consistent with above analysis on the phase
competition in LYCO. However, as we decrease T$_{D}$ further, the
T'-214 phase prefers the (110)-oriented epitaxy on STO or LAST
substrates, which is usually observed in the film preparation of
other cuprate perovskite families. When T$_{D}$=600$\celsius$, the
films are nearly pure (110)-oriented. Only (110) and (220) peaks can
be observed.

Therefore, to acquire the high quality T'-LYCO films, the optimal
deposition temperature should be around 690$\celsius$. The optimal
films are of the highly c-axis oriented T'-214 structure without
detectable T-phase impurity by x-ray diffraction. The typical full
width at half maximum (FWHM) of rocking curves through (006)
reflection of the optimal films is only 0.28\degree, which
confirms the good epitaxial growth of these films.

The surface morphology of typical films, studied by AFM, are shown
in Fig. 3. Although the highly pure c-axis oriented T'-structure in
the LYCO films has been confirmed by x-ray analysis, some randomly
distributed particles on the film surface can be observed. The
similar particles were also observed in the LCCO films grown by
magnetron sputtering, which were considered mainly cuprous oxide
Cu$_{2}$O \cite{lzhao}. Excessive Cu element in the films exists in
the form of CuO$_{x}$ particles, which suggest that there exists
preferential sputtering during the deposition process. The higher
deposition temperature help to crystallize these amorphous oxide
particles and make them detectable by x-ray diffraction, as shown in
Fig. 2(a). According to our experiences in the previous transport
measurements in LCCO film prepared by magnetron sputtering, we
believe that these CuO$_{x}$ particles on the surface  have no
observable effect on the transport properties of the films.

\begin{figure}
\includegraphics[width=0.32\textwidth]{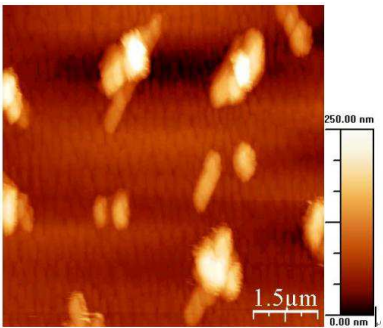}
\caption{Typical AFM image of the LYCO thin film deposited on the
STO substrate at 690$\celsius$.}
\end{figure}

The films in Fig. 4(a) are deposited at higher T$_{D}$, in which
highly insulating T-phase  co-exists as an impurity with
superconducting T'-phase component. The coexistence leads to the
partial superconducting (SC) transition (caused by the T'-phase
component) appended to the whole insulating background (by the
T-phase one). Upon increasing T$_{D}$, the proportion of the T'-214
component reduces and the SC transition weakens, shifting to lower
temperature at the same time  as shown in the inset of Fig. 4(a).

\begin{figure}
\includegraphics[width=0.40\textwidth]{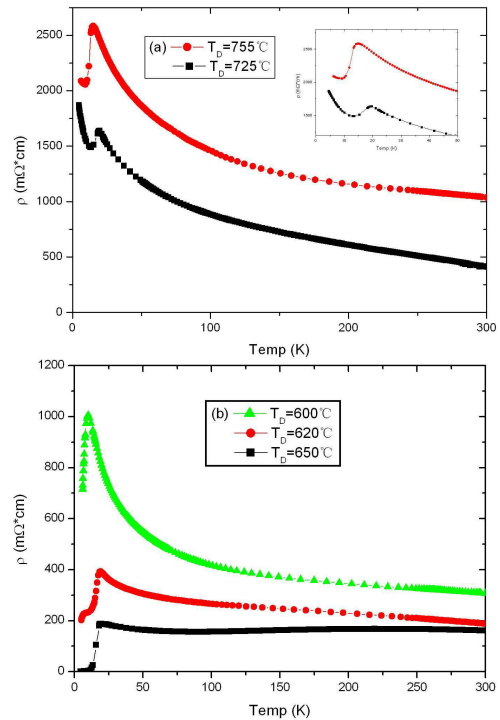}
\caption{The temperature dependence of the resistivity $\rho$ of
films deposited at different T$_D$: (a)755$\celsius$,
725$\celsius$; (b) 650$\celsius$, 620$\celsius$, 600$\celsius$.
The inset in (a) shows superconducting transitions in detail.}
\end{figure}

The films in Fig. 4(b) are deposited at lower T$_{D}$. As T$_{D}$
decreases farther, the (110) component develops from  the c-axis
oriented optimal films, and becomes dominant near
T$_{D}$=600$\celsius$, which has been shown in Fig. 2(e). The (110)
component introduces the contribution of out-of-plane transport to
the whole $\rho-\tt{T}$  behavior of T'-LYCO films. On the other
hand, lower deposition temperature rather strongly hinders the
crystallization process and make deoxygenation  in vacuum more
difficult. At present, we cannot extract $\rho_{c}$ and $\rho_{ab}$
quantitatively from our results because poor crystallization and
grain boundary existing in films deposited at low T$_{D}$. As
T$_{D}$ decreases, the influence of both unremoval of apical oxygen
and grain boundary lead to the larger resistivity above T$_{C}$ and
more incomplete SC transition as well. It is possible that the
semiconducting nature in normal state could arise from the
contribution of out-of-plane resistivity. Similar results are
reported in the study on the anisotropy of
Nd$_{2-x}$Ce$_{x}$CuO$_{4}$(NCCO) through the films with different
orientations \cite{NCCOfilms}.

For the optimal T'-214 films deposited at 690$\celsius$, the temperature dependence
 of the resistivity is shown in Fig.5 (a). The T$_{\texttt{C}0}=$13.5K
 and the width of SC transition less than 1.5K (10-90\% criterion) were acquired. The optimal film
 shows a metallic behavior in the whole range above T$_{C}$ except
 for a slight upturn approaching T$_{C}$ when $\texttt{T} < \texttt{T}_{min}\sim$50K.
 In contrast to the well-known linear temperature dependence in the hole-doped
cuprates like YBCO, the nearly quadratic dependence of $\rho$ in the
normal state is observed. This kind of behavior is similar to those
observed in the electron-doped cuprates around optimal doping, such
as Ln$_{2-x}$Ce$_{x}$CuO$_{4}$ (Ln=La, Pr, Nd, and Sm)\cite{Sawa,
Crusellas, Brinkmann, Greene}. Now most researchers consider it as a
Landau-Fermi liquid behavior due to electron-electron scattering.

\begin{figure}
\includegraphics[width=0.40\textwidth]{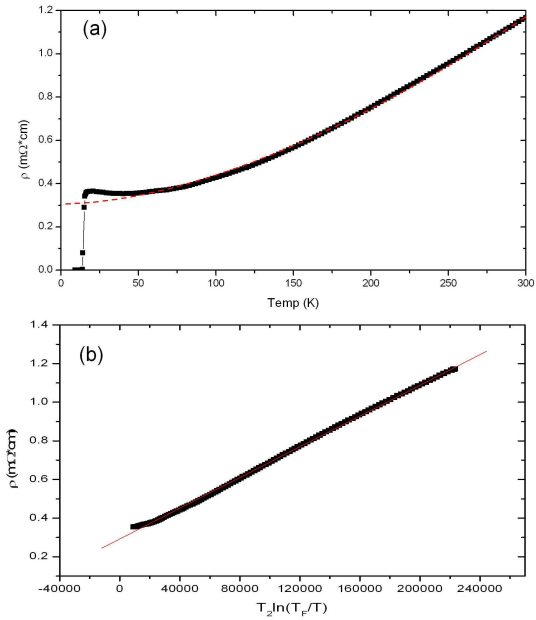}
\caption{Temperature dependence of resistivity for optimal c-axis
oriented T'-LYCO films. (a)$\rho$ vs. T , the dashed line is the
fit to the formula (\ref{fit}); (b)$\rho$ vs.
$T^2\texttt{ln(}T/T_F\texttt{)}$. $T_F=5000$K is used and the
straight in (b) line is a guide to eyes.}
\end{figure}

The rigid formula for two-dimensional Landau-Fermi liquid
\begin{equation} \label{fit}
 \rho(T)=\rho_0+A(T/T_F)^2\texttt{ln}(T/T_F).
\end{equation}
is used to fit the data. Where $T_F$ is the Fermi temperature (we
take $T_F$ =5000K here) and
 $\rho_0$ residual resistivity \cite{Tsuei}. In the normal state the fitting
is quite good above T$_{min}$, which is further confirmed by the
fine linearity in the plot of in Fig. 5(b). The quadratic
temperature dependence with logarithmic correction have been also
observed in other electron-doped cuprates \cite{wangsst}, which
suggest the T'-LYCO may intrinsically be a typical electron-doped
cuprate.

\begin{figure}
\includegraphics[width=0.40\textwidth]{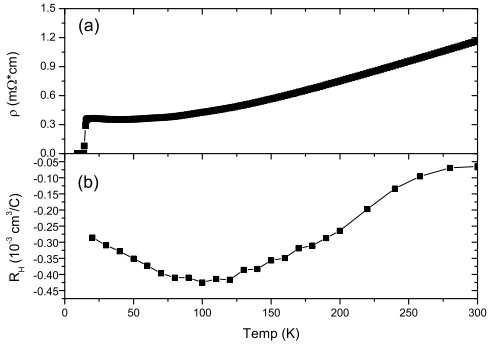}
\caption{Temperature dependence of the resistivity $\rho$ (a) and
Hall coefficient R$_H$ (b) of the optimal film deposited at
T$_D$=690$\celsius$.}
\end{figure}

To clarify the origin of charge carriers, Hall coefficient R$_{H}$
of the optimal films was measured as shown in Fig. 6. The negative
 R$_{H}$  in normal states confirms the intrinsic
electron-doped nature in T'-214 LYCO films. The R$_{H}$  decreases
gradually as temperature cooling from room temperature down to about
100K. Upon decreasing temperature further, a slight upturn toward to
zero occurs. This kind of behavior is quite similar with other
electron-doped cuprates as NCCO \cite{wangprb} and
PCCO\cite{Brinkmann} at slightly less optimal doping level.

\begin{figure}
\includegraphics[width=0.40\textwidth]{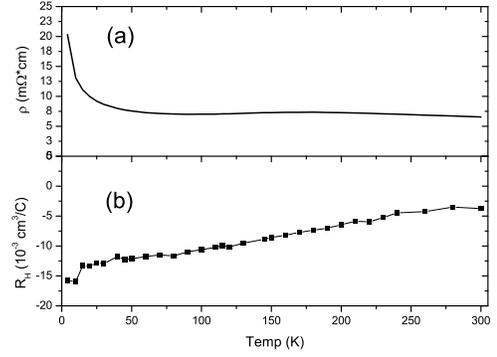}
\caption{Temperature dependence of the resistivity $\rho$ (a) and
Hall coefficient R$_H$ (b) of the LYCO film that was deposited at
T$_D$=690$\celsius$ and annealed at lower vacuum. }
\end{figure}

Till now, we cannot observe any prominent difference in T'-LYCO from
the Ce-doped T'-214 cuprate superconductors. Since Y$^{3+}$ and
La$^{3+}$ are isovalent in T'-LYCO, no net charge can be provided by
doped Y$^{+3}$. The most probable origin of charge carriers is from
oxygen deficiency since the superconductivity of our films can only
be acquired by annealing in high vacuum.

At present, there is no way to determine the oxygen content in thin
films accurately and straightforwardly. To confirm our assumption,
we deposited LYCO films at the optimal temperature (T$_{D}$
=690$\celsius$) but annealing them in lower vacuum (P$_{O_2}\sim
0.6\times 10^{-2}$}Pa) by tuning the slide valve at the front end of
the molecular pump. These films are still of pure T'-214 phase, but
with less oxygen deficiency.

Their resistivity are much larger than the previous optimal films
annealing in high vacuum (P$ <1.0\times 10^{-4}$Pa). At low
temperature the insulating behavior is observed, and no
superconductivity is found down to 4 K as shown in Fig. 7(a). The
Hall coefficient R$_H$ was also investigated as shown in Fig. 7(b).
R$_H$ is negative in the whole temperature range and decreases with
decreasing temperature. The absolute value of R$_H$ is much larger
than that for optimal films, indicating less effective carrier
density and being consistent with larger resistivity. This oxygen
dependence of Hall behavior has been observed in
Nd$_{1.85}$Ce$_{0.15}$CuO$_{4-\delta}$ films by varying oxygen
content \cite{Jiang}. Because the limit of vacuum that the
sputtering chamber can reach is lower than that of the MBE chamber
in the NTT group(P$_{O_2}<10^{-8}$ Torr) \cite{naito, noda}. Tuning
the films to the overdoped region in the phase diagram  cannot be
performed by introducing more oxygen deficiency currently.

\section{CONCLUSIONS}
We have successfully synthesized the superconducting T'-phase LYCO
films by dc magnetron sputtering. The optimal growth conditions are
explored in detail. The appropriate deposition temperature and
succedent \emph{in situ} annealing in high vacuum are the most
crucial to acquire superconducting films with pure c-axis
orientation. The optimal T$_{C}$  of our samples is lower than the
reported MBE-grown films (above 20K) at present. The biggest
restriction to improve T$_{C}$ is the limit of vacuum one can
achieve. Detailed studies on the resistivity and Hall coefficient on
T'-LYCO films, suggesting that this kind of so-called ``non-doped"
superconductor is intrinsically electron-doped cuprates. No exciting
prominent difference in T'-LYCO from the well-known Ce-doped T'-214
cuprate superconductors(as LCCO, NCCO and PCCO) is observed in
experiments. The charge carriers are the most likely from severe
oxygen deficiency. But besides reducing \emph{t} factor to improving
the stability of T'-structure slightly, the role of Y$^{3+}$  ion in
superconducting T'-LYCO is still elusive now. We have prepared
Y-doped T'- Nd$_{2}$CuO$_{4}$ and Pr$_{2}$CuO$_{4}$ in films and
bluk forms, in which there is no need for Y$^{3+}$ ions to stabilize
T'-214 structure. But no superconductivity is observed
\cite{nycopyco}. We suppose that the cooperation of large-sized
La$^{3+}$ ions with smaller Y$^{3+}$ may enable the sufficient
deficiency of oxygen in T'-structure to induce superconductivity.
 Further studies on the T'-LYCO films at various Y-doping levels
 and the in-plane substitution effect by magnetic or non-magnetic
 impurities are currently in process.

\begin{acknowledgments}

The authors would like to thank  G. Y. Wang, T. Wu, B. Guan and X.
G. Luo, for their assistance on the measurements and useful
discussions. This work is supported by the Nature Science Foundation
of China and by the Ministry of Science and Technology of China (973
project No: 2006CB601001), and by the Knowledge Innovation Project
of Chinese Academy of Sciences.

\end{acknowledgments}

\end{document}